\documentclass[modern]{aastex61}

\shorttitle{KIC 9832227 Observed with Vulcan}
\shortauthors{Socia et al.}

\begin{document}

\title{KIC 9832227: Using Vulcan Data to Negate The 2022 Red Nova Merger Prediction}

\author[0000-0002-7434-0863]{Quentin J Socia}
\affiliation{Department of Astronomy, San Diego State University, San Diego, CA 92182, USA}

\author[0000-0003-2381-5301]{William F Welsh}
\affiliation{Department of Astronomy, San Diego State University, San Diego, CA 92182, USA}

\author{Donald R Short}
\affiliation{Department of Astronomy, San Diego State University, San Diego, CA 92182, USA}

\author[0000-0001-9647-2886]{Jerome A Orosz}
\affiliation{Department of Astronomy, San Diego State University, San Diego, CA 92182, USA}

\author{Ronald J Angione}
\affiliation{Department of Astronomy, San Diego State University, San Diego, CA 92182, USA}

\author[0000-0002-6742-4911]{Gur Windmiller}
\affiliation{Department of Astronomy, San Diego State University, San Diego, CA 92182, USA}

\author{Douglas A Caldwell}
\affiliation{SETI Institute, Mountain View, CA 94043, USA}

\author{Natalie M Batalha}
\affiliation{NASA Ames Research Center, Mountain View, CA 94035, USA}

\begin{abstract}

KIC 9832227 is a contact binary whose 11 hr orbital period is rapidly changing. Based on the apparent exponential decay of its period, the two stars were predicted to merge in early 2022 resulting in a rare red nova outburst. Fortunately KIC 9832227 was observed in 2003 as part of the NASA Ames pre-\textit{Kepler} Vulcan Project to search for transiting exoplanets. We find that the Vulcan timing measurement does not agree with the previous exponential decay model. This led us to re-evaluate the other early epoch non-\textit{Kepler} data sets, the Northern Sky Variability Survey (NSVS) and Wide Angle Search for Planets (WASP) survey. We find that the WASP times are in good agreement with the previous prediction, but the NSVS eclipse time differs by nearly an hour. The very large disagreement of the Vulcan and NSVS eclipse times with an exponentially decaying model forces us to reject the merger hypothesis. Although period variations are common in contact binaries, the physical cause of the period changes in KIC 9832227 remains unexplained; a third star scenario is unlikely. This study shows the data collected by the Vulcan photometer to be extremely valuable for extending the baseline for measurements of variable stars in the \textit{Kepler} field.

\end{abstract}


\section{Introduction} \label{sec:intro}

Red novae are a recently discovered class of stellar transients whose change in luminosity is surpassed only by supernovae.  Little is known about red novae, but they are thought to be the result of the merger of the cores in contact binaries. Due to large amounts of gas and dust dispersed by the outburst, the luminosity peaks in the \textit{I}-band \citep{2007Natur.447..458K}. The first confirmed observation was in M85 \citep{2007Natur.447..458K}, though the progenitor to this outburst was unresolved in archival \textit{Hubble Space Telescope} (\textit{HST}) data. V838 Monoceritos was long thought to be an unusual, highly reddened nova eruption, but was later reclassified as a luminous red nova whose pre-outburst spectra suggest a binary star progenitor \citep{2003Natur.422..405B}. The binary nature of red novae progenitors was finally confirmed in Nova Sco 2008 where archival data revealed the precursor was the contact binary system V1309 Scorpii. The progenitor had a $P \approx$ 1.4 day and exhibited an exponentially decaying period \citep{2011A&A...528A.114T}. The merger resulted in a brightening in the \textit{I}-band by 10 mag. Such stellar mergers are common in our galaxy, with a bright red nova outburst event occurring every 10-50 years, and many fainter events from low-mass progenitors happening that cannot be seen \citep{2014MNRAS.443.1319K}. However, a recent search with the OGLE-III survey was unable to identify any merger candidates in the galactic bulge field with rapidly decreasing periods \citep{2017AcA....67..115P}.

KIC 9832227 was originally classified as an RRc Lyrae type pulsator, but using Kepler data \citet{2013arXiv1310.0544K} show it to be an eclipsing (over)contact (W UMa) binary system . A subsequent study by \citet{2017ApJ...840....1M} shows significant changes in its $\approx$ 11 hr period using eclipse times spanning almost two decades fit with an exponentially decaying function of the same form as V1309 Scorpii ($P \propto \exp(1/(t_{0}-t))$). The parameter $t_0$ is the time of merging, predicted to be in the year 2022. Such a merger should result in a red nova, reaching an apparent visual magnitude of $\sim$2 \citep{2017ApJ...840....1M}. Thus, KIC 9832227 could provide us with a rare opportunity to study a red nova progenitor in detail before outburst. To test and better constrain the decaying period hypothesis, we used archival Vulcan data from 2003 \citep{2004IAUS..202...69C} to add another point to the Observed minus Computed (\textit{O-C}) diagram at an early epoch. We also provide our own observations from the summer of 2017.

\section{The Vulcan Photometer} \label{sec:vulcan}

The Vulcan Photometer was a 10 cm aperture ground-based instrument at the Lick Observatory designed to detect Jovian-size planets around Sun-like stars (\citealt{2001PASP..113..439B}; \citealt{2002ApJ...564..495J}). It observed about 20,000 stars brighter than 13th magnitude in a 49 deg$^2$ field of view for about 90 days in the summer of 2003. The Vulcan project discovered numerous variable stars, many of which are eclipsing binaries or multi-star systems \citep{2004IAUS..202...69C}.  The sky surveyed by Vulcan shares a substantial overlap with the Kepler field making the survey particularly valuable.

We de-archived and recalibrated the Vulcan observations of KIC 9832227. Eclipse times are crucial to this study, so to confirm the accuracy of the Vulcan times, we measured the eclipse times of KIC 9592855, a binary system that has an ephemeris measured to a high precision and shows no evidence for a variable period \citep{2017ApJ...851...39G}. We find the Vulcan timing to be within 20 s of the \textit{Kepler} eclipse times projected back to the 2003 Vulcan epoch. This is within 1$\sigma$ of the propagated uncertainty in the ephemeris. To further confirm the stable period of KIC 9592855, we measured eclipse times from the Wide Angle Search for Planets (WASP; \citealt{2010A&A...520L..10B}) survey data from 2007 and 2008 to be within 1$\sigma$ (45 s) of the expected time. (All times were converted from UTC to BJD using the Ohio State University UTC2BJD Time Utility\footnote{http://astroutils.astronomy.ohio-state.edu/time/}; \citealt{2010PASP..122..935E}.)

\section{Observations} \label{sec:obs}

\textit{Kepler} data from Quarters\footnote{Note that the Molnar et al.\ study omitted Quarters 4, 12, and 13. These Quarters are also missing from the Kepler Eclipsing Binary Catalog V3.} Q0 through Q17 (\citealt{2011AJ....141...83P}; \citealt{2016AJ....151...68K}) were downloaded from MAST and obvious bad data were eliminated. To remove any remaining systematic calibration errors, continuous segments of  the light curve were detrended by dividing by a low order polynomial fit. Each of these segments was then concatenated into one normalized light curve. To measure the Kepler primary eclipse times, an iterative method was employed. Each iteration began with a set of estimated eclipse times which were used to produce a stacked light curve two-cycles long in phase (i.e. phases 0.0-2.0, with primary eclipse at phase 0.0, 1.0, and 2.0). A piecewise cubic Hermite spline (PCHS) was then fit to the stacked data to construct a template. [The PCHS used 35 nodes, with the nodes at the primary eclipse minima (middle and end points) forced to have zero derivative to minimize any skew in the fit.] The template was then shifted across a small neighborhood of time surrounding each of the 2857 primary eclipses, and the chi-square is used to give refined estimates of the eclipse times with uncertainties. The process was then iterated until convergence. The final set of times was then used to derive a linear ephemeris, resulting in $T_{0}$ = 2454953.949183 $\pm \ 9\times 10^{-6}$ BJD and $P_{Kep}$ = 0.457948557 $\pm \ 5\times 10^{-9}$ days. Figure 1 shows the \textit{O-C} times for these eclipses. The \textit{Kepler} \textit{O-C} shows a curious feature that oscillates for the first half of the dataset with a period of about 100 days, likely due to starspots creating a beat period with the binary orbit. This corresponds to a starspot period of $P_{\mathrm{starspot}} = 0.46$ days. Variations like these have been previously shown to be caused by starspots, supported by the anticorrelated nature of primary and secondary eclipse timings for this system  \citep{2013ApJ...774...81T}. These variations of up to 10 minutes are present, setting the limit on the accuracy (not precision) of any individual eclipse timing measurement.

The Northern Sky Variability Survey (NSVS; \citealt{2004AJ....127.2436W}) was queried and KIC 9832227 appears twice, as object 5597755 and 5620022 in two different camera fields. The two were then combined into one light curve.\footnote{No corrective offset was applied to either object. Following \citet{2017ApJ...840....1M}, we offset observation set 5620022 by -0.062 mag before combining the two into one light curve. Measurements with and without the magnitude offset are within 1$\sigma$. Due to an inconsistent sampling for each field, finding an optimal offset is nontrivial and beyond the scope of this Letter.} The NSVS times were converted from MJD (MJD $\equiv$ JD - 2450000.5) to JD, then to BJD \citep{2010PASP..122..935E}. Two data sets from the WASP survey were obtained, from 2007 and 2008, with both seasons consisting of about 90 days of observation. Two other seasons were determined to be too incomplete to measure reliable eclipse times. Times were converted from HJD to BJD using the HJD2BJD utility \citep{2010PASP..122..935E}. Observations were also made with the Mount Laguna Observatory (MLO) 40 inch telescope in 2017 on the nights of June 2, 3, 30, July 1, and September 25. Exposures ranged from 15 to 20 s and were made in the Johnson-Cousins R filter. The data were reduced using AstroImageJ (AIJ; \citealt{2017AJ....153...77C}), removing the usual CCD bias, performing flat-field division, and differential photometry using comparison stars in the 14.5 square arcmin image field. AIJ utilizes the Ohio State UTC2BJD calculator to convert JD to BJD times. Figure 2 shows the Vulcan, \textit{Kepler}, and MLO phase-folded data. Notice the clear offset in the eclipse minima. The eclipse times were measured by phase folding the light curve from each data set on the \textit{Kepler} ephemeris ($P_{Kep}$ = 0.457948557 days), then shifting the \textit{Kepler} template to best match the observations. The BJD$_0$ time listed by the Kepler Eclipsing Binary Catalog \citep{2016AJ....151...68K} was used to identify the primary eclipse; this is consistent with the radial velocity determination of the primary star \citep{2017ApJ...840....1M}. Table 1 lists the measured primary eclipse times.

\section{Analysis and Discussion} \label{sec:an}

To our surprise, the measurements of the eclipse times in the Vulcan data do not match the exponentially decaying prediction of Molnar et al. Furthermore, our NSVS measurement differs by +1.01 hr from the Molnar et al.\ reported eclipse time \citep{2017ApJ...840....1M}. Based on the evidence from the Vulcan data, we speculate that the very large discrepancy with the NSVS timing measurement may be caused by a cycle count error and an MJD to BJD conversion error. Since the period of the binary is 10.99 hr, if the 0.5 day offset from the MJD to JD conversion were omitted, and the eclipses were one cycle off, there would be a difference in the eclipse time of 0.99 hr. The red X in Figure 3 is the hypothetical data point with these two factors included. The X matches the exponential decay curve remarkably well, lending credibility to our hypothesis. Furthermore, \citet{2004AJ....127.2436W} defines in their Table 6 MJD $\equiv$ JD - 2,400,000; this unfortunately is different from the preprint version of the paper\footnote{arXiv:astro-ph/0401217} which includes the standard half-day offset. In addition, L.A. Molnar (2018, private communication) pointed out that from Los Alamos, KIC 9832227 would not be above the horizon at the time reported in \citet{2004AJ....127.2436W}, thereby clinching the MJD timing error hypothesis. We henceforth ignore the fictitious datum given by the red X. Using all valid data points, we determined a new linear ephemeris for the binary to be $T_0$ = 2454953.48885 BJD $\pm \ 9\times 10^{-5}$ days, and $P$ = 0.45794896 days $\pm \ 5\times 10^{-8}$ days. This ephemeris is used as the ``C" in the \textit{O-C} of Figure 4.

The \textit{O-C} does show real changes in period. In particular, the MLO eclipses occur 40 minutes early with respect to the calculated \textit{Kepler} ephemeris. Such a change is not unusual for contact binaries: mass transfer, magnetic braking, or magnetic cycles are plausible explanations (\citealt{1992ApJ...385..621A}; \citealt{1994A&A...282..775K}), though magnetic braking would only cause a decrease in period and would not explain the increases. A majority of contact binary stars are found to have changes in their periods with $\dot{P}$ evenly distributed around zero and up to $\pm \ 2.3\times 10^{-7}$ day year$^{-1}$, indicating that decreasing periods are not favored \citep{2006AcA....56..253K}. Systems like AB Andromedae \citep{2008Ap&SS.317...71H}, AH Virgo, and V566 Ophiuchus vary in their eclipse times by up to 2 hr. V502 Ophiuchus \citep{1994A&A...282..775K} is a contact binary with a period of 0.4533925 days that shows a 40 minute change in its \textit{O-C} times, quite similar to KIC 9832227. We therefore conclude that since KIC 9832227 does not exhibit the same type of period decay as V1309 Scorpii, it is very unlikely it will merge in 2022.

Contact binaries commonly have tertiary components ($\sim$60\% occurrence rate), as a third star is likely necessary to remove angular momentum allowing the binary to tighten into a contact system (\citealt{2006AJ....132..650D}; \citealt{2006AJ....131.2986P}; \citealt{2007AJ....134.2353R}; \citealt{2013ApJ...768...33R}; \citealt{2014AJ....147...87T}; \citealt{2017AcA....67..115P}). The occurrence rate of triple systems in the \textit{Kepler} close binary sample is estimated by \citet{2014AJ....147...45C} to be about 10\% and by \citet{2013ApJ...768...33R} to be 20\% . For longer period binaries (P $>$ 1 day), the occurrence of triple star systems is 15\%-20\% \citep{2015ASPC..496...55O}. It is therefore quite possible that the \textit{O-C} variation is induced by a third star. We fit the eclipse timing variations using the Python package \textit{lmfit} \citep{2016ascl.soft06014N}, a Levenberg-Marquardt nonlinear least-squares fitting algorithm, to find orbital parameters for a possible third star. The mass of the third body was constrained by the combined masses of the primary and secondary star, from measurements made by \citet{2017ApJ...840....1M}, which is 1.714 $M_\odot$. Also, the hypothetical third star does not appear in the spectra \citep{2017ApJ...840....1M}, which puts an upper limit to its mass. \citet{2017ApJ...840....1M} rule out a hypothetical star of mass $\geq$ 0.8$M_\odot$. Three different mass solutions are shown in Figure 4, none of which provide a good match to the data. We used \textit{emcee}, a Markov chain Monte Carlo ensemble sampler \citep{2013PASP..125..306F}, to set a lower limit on the period of a third star of 7200 days, or about 20 years. This period is, of course, suspicious as it is roughly the duration of our observations. The lower limit of the third mass using a fixed 7200 day period is about 0.7 $M_\odot$, which is uncomfortably high considering it is not detected in the spectra.

Our original intent was to strengthen the very exciting red nova prediction and to better characterize the decay of the suspected progenitor binary. Our results cast doubt on the prediction that the stars will merge in 2022, as the data do not follow the exponential decay at early epochs. The \textit{O-C} does show real changes in period, but our most favored explanation is that these are merely period wanderings common in many contact binaries. This investigation, however, does demonstrate the substantial value of the Vulcan project data for extending the baseline of \textit{Kepler} eclipse timing measurements.

We thank Dr.\ Larry Molnar for helping improve this paper and especially for confirming the MJD offset error in the NSVS data. We acknowledge the support from NASA via grants NNX13AI76G (OSS) and NNX14AB91G (Kepler PSP), and from the NSF via grant AST-1617004. This research made use of the NExScI Exoplanet Archive\footnote{https://exoplanetarchive.ipac.caltech.edu/} and the Kepler Eclipsing Binary Catalog V3 hosted at Villanova University.\footnote{http://keplerebs.villanova.edu/} This publication makes use of the data from the Northern Sky Variability Survey created jointly by the Los Alamos National Laboratory and University of Michigan. The NSVS was funded by the U.S. Department of Energy, the National Aeronautics and Space Administration, and the National Science Foundation.

\facilities{SuperWASP, Kepler, MLO}

\software{AstroImageJ \citep{2017AJ....153...77C}, emcee \citep{2013PASP..125..306F}, lmfit \citep{2016ascl.soft06014N}}

\bibliographystyle{apj}
\bibliography{ref}

\begin{figure}[ht]
  \centering
    \includegraphics[width=1\textwidth]{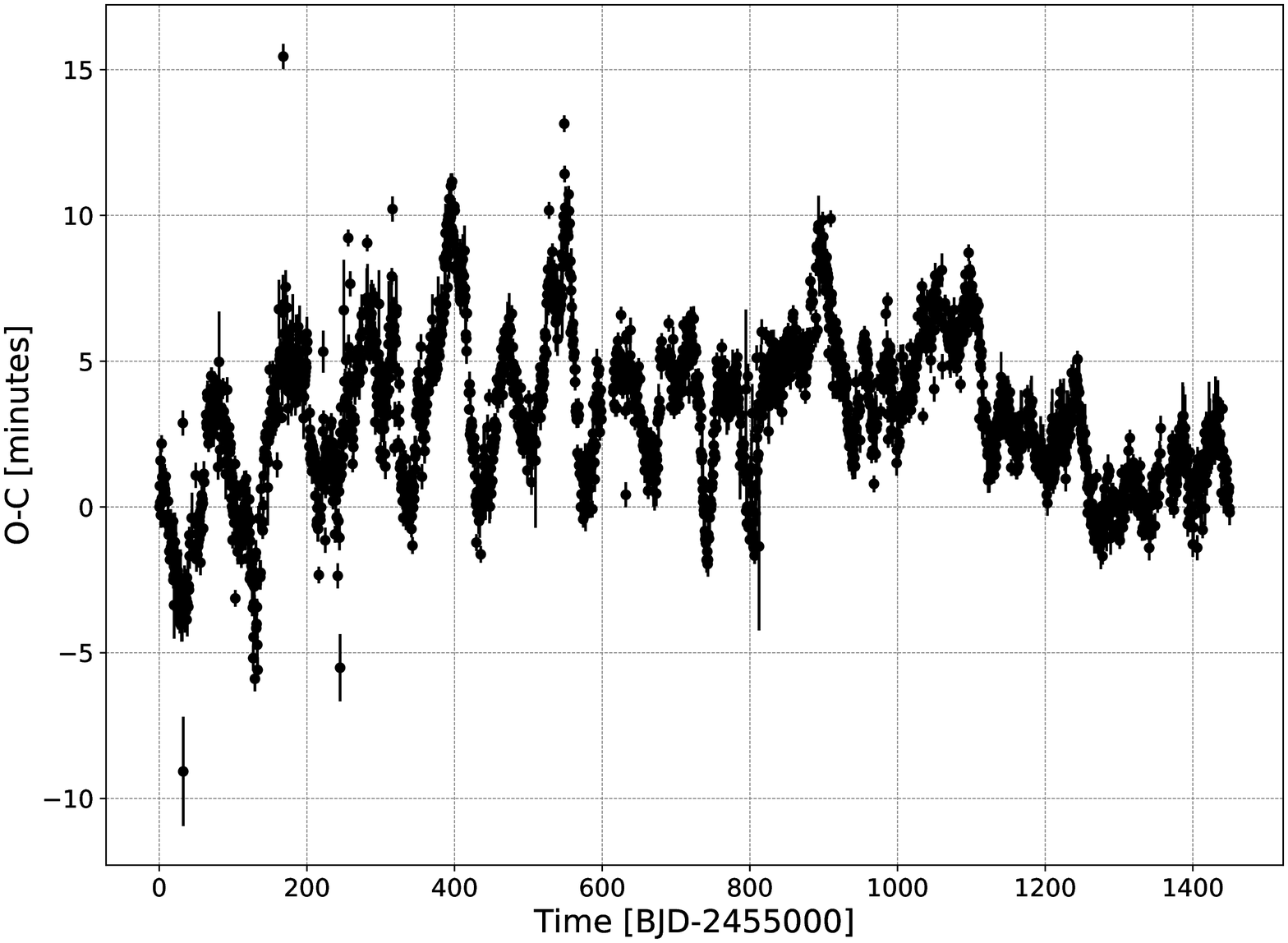}
  \caption{The \textit{O-C} diagram of \textit{Kepler} primary eclipse timings of KIC 9832227. A quasi-periodic feature in the early portion of the \textit{O-C} and a slight long-term downward curvature are present. The oscillation is possibly caused by starspots which would have a rotation period of 0.460 days. This is slightly longer than the 0.458 day orbital period, suggesting transient spots at different latitudes on a differentially rotating star cause the short-term changes in the \textit{O-C}.}
\end{figure}

\begin{figure}[ht]
  \centering
    \includegraphics[width=1\textwidth]{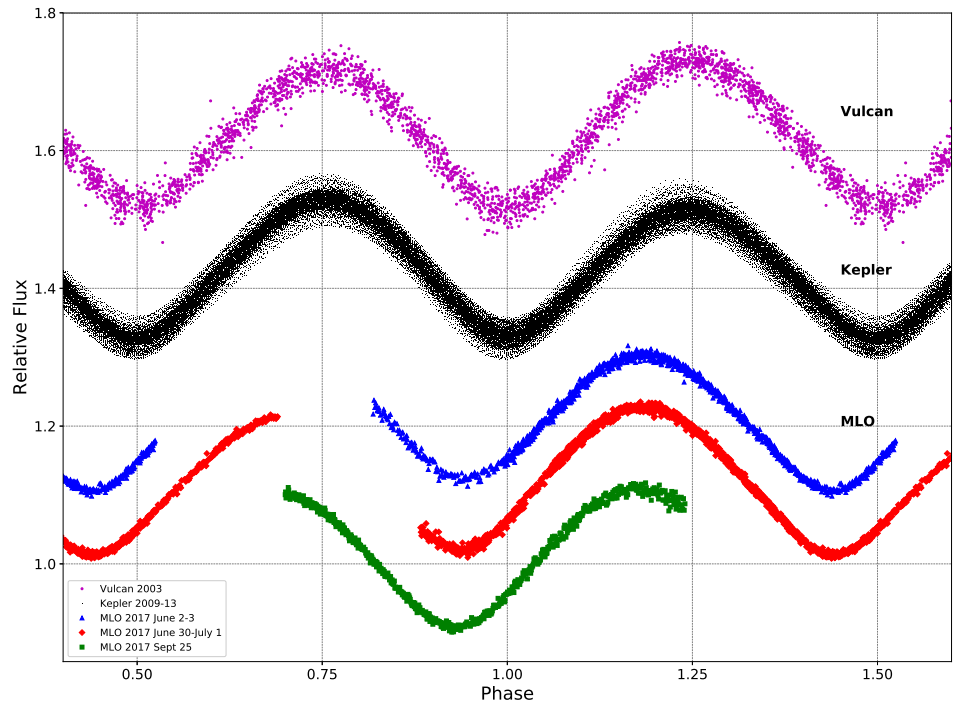}
  \caption{Normalized and detrended Vulcan, \textit{Kepler}, and MLO light curves phase folded on our ephemeris determined from \textit{Kepler} data. Each data set is offset in flux for comparison.}
\end{figure}

\begin{figure}[ht]
  \centering
    \includegraphics[width=1\textwidth]{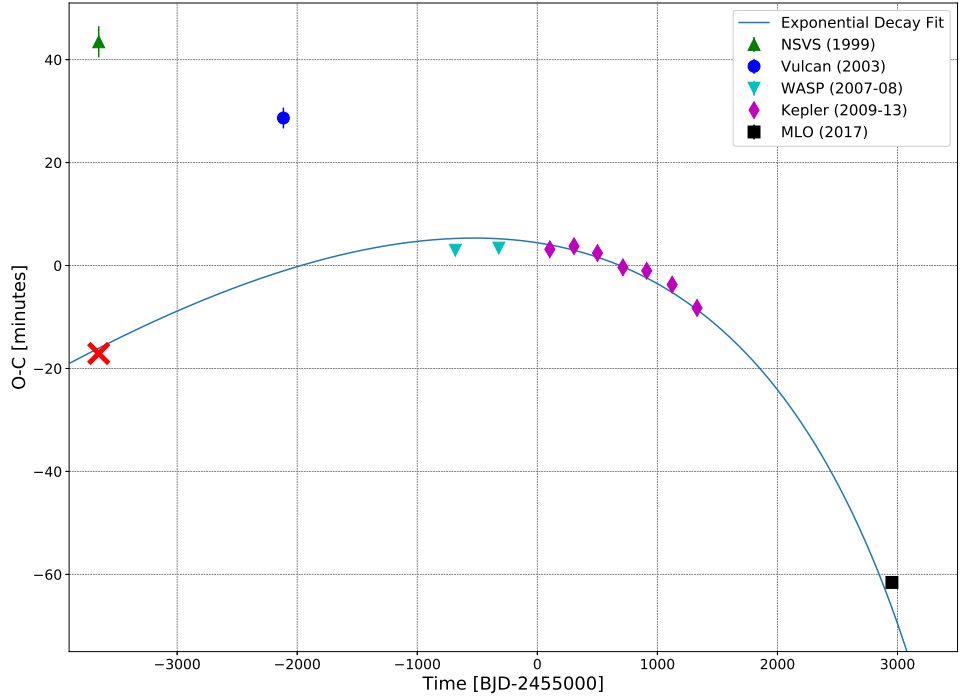}
  \caption{The \textit{O-C} diagram of KIC 9832227 with a reference period of $P_{\mathrm{ref}} = 0.4579515$ days. The blue curve is a recreation of the exponentially decaying fit from \cite{2017ApJ...840....1M} with the same reference period. The red 'X' on the left of the figure is the NSVS datum with a 1 cycle count and MJD +0.5 day error, creating a plausible, but misleading datum. Our measured value, the green triangle, is at +42 minutes, which is about 1.01 hr later. To prevent the 2857 eclipses in the \textit{Kepler} data from completely dominating our fit, the \textit{O-C} values were cast into 7 bins of approximately 200 days in length.}
\end{figure}

\begin{figure}[ht]
  \centering
    \includegraphics[width=1\textwidth]{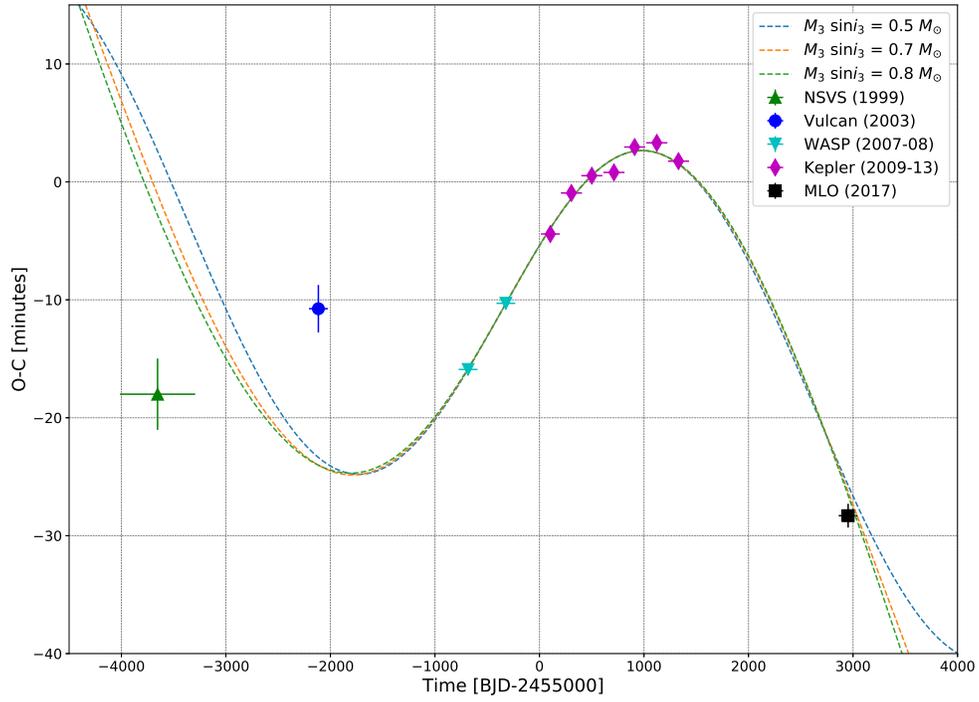}
  \caption{Revised \textit{O-C} diagram using all of our eclipse times. The horizontal line on each datum represents the duration of each data set. Three tertiary star fits to the \textit{O-C} data are shown, none of which agree well with NSVS or Vulcan eclipse times. }
\end{figure}

\clearpage

\begin{table}[ht]
	\centering
	\caption{KIC 9832227 Primary Eclipse Times}
	\label{my-label}
	\begin{tabular}{lll}
		\hline
        \hline
		Data Origin (Year) & Eclipse Time (BJD) &  Uncertainty (days) \\ \hline 
		NSVS (1999) &  2451299.9802 & $\pm$0.0021 \\
		Vulcan (2003) &  2452838.2290 & $\pm$0.0014 \\ 
        WASP (2007) &  2454270.6834 & $\pm$0.0002 \\
        WASP (2008) &  2454632.4654 & $\pm$0.0002 \\
        \textit{Kepler} (2009-2013) &  2454953.9492 & $\pm$0.0002 \\
        MLO (2017) &  2457906.7735 & $\pm$0.0003 \\ \hline
	\end{tabular}
\end{table}

\end{document}